\documentstyle[preprint,aps,epsfig]{revtex}
\draft
\begin{document}
\title{The nuclear effects in $g_{1^3\rm{He}}$ 
and  the Bjorken sum rule for A=3}
\author{L.Frankfurt$^{*}$, V. Guzey$^{**}$, M. Strikman$^{**}$}
\address{$^{*}$School of Physics And Astronomy, Raymond and Beverly Sackler 
Faculty of Exact Science\\
 Tel Aviv University, Ramat Aviv 69978, Israel\\
$^{**}$Department of Physics,
The Pennsylvania State University, University Park, PA 16802}
\maketitle

\begin{abstract}
The Bjorken sum rules for the $A=3$ and  $A=1$
are used as a guide to estimate nuclear effects in 
extracting $g_{1n}(x,Q^{2})$ from the 
$\vec{e}\,\vec{^3\rm{He}}$ data.
We estimate that the
combination of the spin depolarization, the nonnucleonic degrees of freedom
in the $A=3$ system
and nuclear shadowing is likely to reduce 
$g_{1^3\rm{He}}(x\le 0.05)$ by $\sim 15 \%$
while 
a significant
enhancement of  the structure functions $g_{1^3\rm{He}}, g_{1^3\rm{H}}$
at $x \sim 0.1$
is predicted.
\end{abstract}


\narrowtext
Over the last decade a series of experiments has been performed aiming at 
measuring the polarized structure 
functions of protons and neutrons, cf. \cite{EK1}. The
primary motivation was to check  the Bjorken sum rule.
Recently it has been emphasized  that high precision 
measurements of this sum rule may allow for the accurate determination of 
$\alpha_s(Q^2)$ \cite{EK}.
The measurement  of $g_{1n}(x)$ involves 
necessarily
nuclear targets. 
Several experiments using polarized $^2$H  and $^3$He
targets have been performed and  several more are in progress.
The  advantage of the $^3$He target over the $^2$H 
target is that in the  
first approximation only the neutron is polarized, 
so that the contribution of the much larger proton structure function
$g_{1p}(x,Q^2)$ is small \cite{Woloshin}.

High precision nonrelativistic calculations of the 
$^3$He wave function using realistic nuclear potentials 
are now available, cf. \cite{Friar}. 
They have been applied to analyze the polarized e--$^3$He scattering using the 
convolution models,  where nonnucleonic degrees of 
freedom in nuclei and nuclear shadowing are neglected \cite{Ciofi,Sauer}.

The general conclusion is that, similar to the $^2$H case \cite{FS83},
the major effect of nuclear structure for $x \le 0.5$
is the depolarization of nucleons 
in nuclei due to the presence of the
higher partial waves.  Fermi motion effects 
do not produce any noticeable 
$x$ dependence up to $x \sim 0.5$ \cite{Ciofi,Sauer}.
To avoid dealing with small corrections due  to the $\sim 2 \%$ polarization
of protons in $^3$He
it is convenient to consider the nonsinglet polarized structure functions:
$$g_{1N}^{n.s.}(x,Q^2)\equiv g_{1p}(x,Q^2)-g_{1n}(x,Q^2),$$
and 
$$g_{1,A=3}^{n.s.}(x,Q^2)\equiv g_{1^3\rm{H}}(x,Q^2)-g_{1^3\rm{He}}
(x,Q^2).$$ One finds \cite{Ciofi,Sauer}
\begin{equation}
g_{1,A=3}^{n.s.}(x,Q^2)= (P_S-{1\over 3}P_{S\prime}+{1\over 3}P_D)
g_{1N}^{n.s.}(x,Q^2),
\end{equation}
for $x \le 0.5$. Here $P_S$, $P_{S\prime}$ and $P_D$
are the probabilities of the corresponding components of the neutron 
wave function in $^3$He.

For  the ratio of the Bjorken sum rule for $A=3$ to $A=1$
within the discussed above impulse approximation the  corrections which
 are proportional to $\alpha_s^n(Q^2)$ cancel out and one obtains:
\begin{equation}
R = 
\frac{\int^{1}_{0}[g_{1}^{^{3}He}(x,Q^2)-g_{1}^{^{3}H}(x,Q^2)]dx}
{\int^{1}_{0}[g_{1}^{n}(x,Q^2)
- g_{1}^{p}(x,Q^2)]dx} = \frac{G_{A}(^3\rm{H})}{G_{A}(n)},
\label{Bj}
\end{equation}
 independent of $Q^2$, where  we have ignored the higher twist effects.
$G_{A}$ is the axial coupling constant for $\beta$ decay of the nucleus A.
Comparing eqs. (1) and (2) we find
\begin{equation}
R=P_S-{1\over 3}P_{S\prime}+{1\over 3}P_D.
\end{equation}
This is perfectly consistent with the expression for $G_A(^3\rm{H})$ derived
by Blatt back in 1952 \cite{Blatt}:
\begin{equation}
G_{A}(^3 {\rm H})=(P_{S}-{1\over 3}P_{S\prime}+{1\over 3}P_D)G_{A}(n).
\label{ga}
\end{equation}

The problem however is   that  relation (\ref{ga}) is known to be violated
experimentally rather significantly. Indeed, 
realistic 3-nucleon models of $^3$He and $^3$H give\cite{Friar}:
\begin{equation}
P_S-{1\over 3}P_{S\prime}+{1\over 3}P_D=1-(0.0785 \pm 0.0060).
\end{equation}
Combining the most recent experimental data on 
$G_{A}(^3{\rm H})/G_{V}(^3{ \rm H})$ for  tritium $\beta$-decay \cite{Budick} 
(the data is in 
good agreement with the previous data \cite{ga}) with 
the value of $G_{A}(n)/G_{V}(n)$ from \cite{Particle})
 we 
obtain
\begin{equation}
\frac{G_{A}(^{3}{\rm H})}{G_{A}(n)}
 = 1-(0.0366 \pm 
0.0030).
\end{equation}

Hence we conclude that {\it the use of the 
convolution model,
combined with the 3-nucleon description of $A=3$ nucleon system, leads to 
a $\sim 4\%$ violation of the Bjorken sum rule for the scattering of the $A=3$ 
systems}. 
This is consistent with the  general expectation 
that noticeable nonnucleonic degrees of freedom should be present in the 
$A=3$ systems.

Nuclear effects for the Bjorken sum rule were 
first discussed by Close et al \cite{Close} and by Kaptari 
and Umnikov \cite{kaptari}. In particular it was pointed out in 
Ref.\cite{kaptari} 
that
convolution models and
three nucleon description of $A=3$ system lead 
to results for $g_{1 ~A=3}$ inconsistent with the  Bjorken sum rule,
though they did not notice consistency of eqs. (3) and (4) which is of
 importance for our subsequent analysis.
This observation was left unnoticed in Refs.\cite{Ciofi,Sauer}
and in all analyses of the experimental data.  

The importance of the $\Delta$-isobar and meson exchange currents 
for a quantitative explanation of the 
value of  $G_{A}(^3 {\rm H})/G_{A}(n)$ is discussed in  literature
for a long time, see e.g. Ref.\cite{EW}.
The recent theoretical analyses of $G_{A}(^3\rm{H})$
 \cite{delta,delta1} confirm the conclusion of Ref.\cite{EW} that the dominant
contribution originates from the 
admixture of $\Delta$-isobars  in $^{3}$He and in $^{3}$H. They 
lead to a 
value of $G_{A}(^3 { \rm H})/G_{A}(n)$ consistent with eq. (6). For 
example, Ref. 
\cite{delta} gives $G_{A}(^3 {\rm H})/G_{A}(n)=1-0.0378 \pm 0.002$.
This implies 
that  the major correction to the impulse approximation calculation of 
$G_{A}(A=3)$ is due to $\Delta \rightarrow N$ transitions. Thus a 
natural mechanism for resolving  the  discrepancy
between the Bjorken sum rule 
for $A=3$ and for $A=1$ targets 
which is
present in the impulse approximation, is  the necessity to account  
for  the nondiagonal transitions 
$\gamma^*N \rightarrow \gamma^*\Delta$. No theoretical investigations of 
this structure function have been done as yet.  For the simple case of 
$g_{1n}^{n.s}$ one can expect the same low $x$
behavior for this structure function as for the diagonal
transitions since 
Regge trajectories 
with rather close value of intercept
couple in this case.
Based on $SU(6)$ symmetry, for average $x\sim 0.2\div 0.3$ we can expect 
a behavior similar to the diagonal nonsinglet matrix elements. 
Consequently, we can estimate that the contribution of the
$\gamma^*N \rightarrow \gamma^*\Delta$
transition to $g_{1,A=3}^{n.s.}$ leads to a change in the ratio
${g_{1,A=3}^{n.s.}(x,Q^2)\over g_{1N}^{n.s.}(x,Q^2)}$ for $x \le 0.5$ from  
1-(0.0785 $\pm$ 0.0060) to  $G_{A}(^3 { \rm H})/G_{A}(n)= 1-(0.0366 \pm 
  0.0030)$.
Moreover, treating the $\Delta$-admixture as a perturbation we observe
that main contribution to $g_1$ should originate in the
$^3$He case from the $n \rightarrow \Delta^0$ nondiagonal transition
and in the $^3$H case from the $p \rightarrow \Delta^+$
 nondiagonal transition. In the $SU(6)$ limit, which seems reasonable 
at least for the valence quark contribution \footnote{Note that the SMC
semiinclusive data \protect\cite{SMC} seem to indicate that contribution of 
the sea to
$g_1$ is small down to $x \sim 0.01$.}
\begin{equation}
{g_{1~~n \rightarrow \Delta^0}(x,Q^2)\over g_{1n}(x,Q^2)}=
{g_{1~~p \rightarrow \Delta^0}(x,Q^2)\over g_{1p}(x,Q^2)}.
\end{equation}
Hence up to a small correction due to the contribution of $g_{1p}(x,Q^2)$, the 
combined effect of nucleon depolarization and nondiagonal contributions
is approximately 
the same for  ${g_{1,^3He}(x,Q^2)\over g_{1n}(x,Q^2)}$ and for 
${g_{1,A=3}^{n.s.}(x,Q^2)\over g_{1N}^{n.s.}(x,Q^2)}$. 
We can write  in this approximation
\begin{equation}
g_{1^3He}(x,Q^2)={G_A(^3H)\over G_A(n)}g_{1n}(x,Q^2)+2 p_p
(g_{1p}(x,Q^2)+g_{1n}(x,Q^2)),
\end{equation}
\begin{equation}
g_{1^3H}(x,Q^2)={G_A(^3H)\over G_A(n)}g_{1p}(x,Q^2)+2 p_p
(g_{1p}(x,Q^2)+g_{1n}(x,Q^2)),
\end{equation}
where $p_p \approx -2.8\%$ is polarization of a proton in $^3$He.
We neglect here contribution of $\Delta^+\rightarrow p$ nondiagonal
terms since they effectively merely result in the  renormalization of 
 $p_p$ by a factor
 $\sim {G_A(^3H)\over G_A(n)}$.
Experimentally,  for  small $x \le 0.1$
$$\left|g_{1p}(x,Q^2)+g_{1n}(x,Q^2)\right|\ll \left|g_{1n}(x,Q^2)\right|\,
 \left|g_{1n}(x,Q^2)\right|$$ 
and hence the last term  for these x is 
a very small correction.

Let us calculate nuclear effects specific for the small x physics.
At small $x$, when  the coherence length 
$l={1 \over 2 m_Nx}$ far exceeds the nucleus radius, the virtual 
photon converts to a quark-gluon configuration $h$ well before the target.
In the case of nucleon targets this leads to diffraction in deep inelastic 
scattering which has recently been observed at HERA. For the nuclear targets
this leads to the   shadowing phenomenon, for review see \cite{FS88}. 
Currently nuclear shadowing
in the leading twist
is   observed  experimentally for the sea 
quark distribution, for
the
recent review see \cite{Arneodo}. There is indirect 
evidence for the 
presence of this phenomenon for valence quarks \cite{FLS}. 
The presence of gluon shadowing was recently reported based on
the
analysis of the scaling violation of the
 $F_{2Sn}(x,Q^2)/F_{2C}(x,Q^2)$ ratio \cite{Pirner}.

The phenomenon of shadowing reflects the presence of quark-gluon
configurations in $\gamma^*$ which can interact with cross sections
comparable to that of hadrons. A quantitative description of nuclear shadowing 
phenomenon in deep inelastic scattering was developed in the color 
screening models \cite{FS88,FLS,BL,NZ,PRW}, where   
$\gamma^*$ converts to a quark-gluon state $h$ which interacts with 
the nuclear target   via multiple color singlet exchanges.
The effect of shadowing is determined in these models by the average
value of the ratio 
$\sigma_{eff}={\left<\sigma^2\right>\over \left<\sigma\right>}$, where
averaging is taken over different strengths of interaction, that is, over 
different quark-gluon
configurations involved in the transition
$\gamma^* \rightarrow "hadron~state"$. Numerical analyses of
nuclear shadowing for $A \ge 12$ give $\sigma_{eff} \sim 17 ~mb$.
Similar number follows from the estimate based on the generalization  of the 
optical theorem, cf. \cite{MP78} to the diffractive processes
\begin{equation}
\sigma_{eff}= {16 \pi 
{d \sigma(\gamma^*+p \rightarrow X+p)\over dt}_{\left|t=0 \right.}\over 
\sigma_{tot}(\gamma^*+p)}.
\end{equation}
As soon as this parameter is fixed all 
color singlet models give very similar results for 
$x \ll {1 \over 4m_NR_A}$, for a recent  discussion and refs. 
see \cite{Paris}.
We will use this model in the following  analysis.

It follows from the formulae of the Glauber approximation that for
the case of  cross sections which constitute  a small fraction of the 
total cross section, the  shadowing effects 
should be larger. Several 
examples include shadowing in the parity violating $\vec {p} A$
scattering  \cite{FS81} and shadowing for valence quarks \cite{FS88}.
The  underlying physics is quite  simple. Let us consider 
scattering off a heavy nucleus 
in which one nucleon is polarized.  If this nucleon is at a 
small impact parameter the  optical density is high and the cross
 section of the 
interaction is not sensitive to its polarization. Hence the
cross sections for two polarizations would
differ due to large impact parameters 
only, and therefore shadowing is larger in this case than in the 
case of the total   cross section. Consequently we expect an enhancement 
of the contribution due to the nuclear shadowing effect to $g_{1A=3}^{n.s}$ as 
compared to $F_{2\,^3\rm{He}}$. 

To calculate shadowing for the case of $\vec{e}\vec{^{\,3}\,\rm{He}}$ 
scattering for $Q^2 \sim Q^2_0 \sim ~few\ GeV^2$ we can
consider the difference in the cross sections for the scattering 
of $\gamma^*$ with a given helicity (we will not write it explicitly) off 
$ \vec{^3\rm{He}}$ with helicities $\pm 1/2$ which we will denote $\pm$.
(For larger $Q^2$ the scaling violation for $F_{2A}(x,Q^2)$, $g_{1A}(x,Q^2)$ 
  can be accounted for using QCD evolution 
equation.) The cross section can be written in a symbolic form as 
\begin{equation}
\sigma_{\gamma^* \vec{^3He}_{\pm} }=\sum_{h}\Big|
\left<\gamma^*|h\right>\Big|^2\,\sigma (h^3\vec{He}_{\pm})
\end{equation}
We substituted the integral over the hadronic state by its value at an average 
point that has an interaction with a nucleon $\sigma_{eff}$, and 
 mass of the state $h$ is 
$M^2=Q^2$ \cite{FS88}.

For simplicity we consider the model where all nucleons in the nucleus 
of $^3$He  are in the $S-$state and hence only the neutron is polarized.
However we expect that nuclear shadowing effects should lead to a universal
factor weakly dependent on the form of the wave function of the nucleus.
To calculate $\sigma(h^3\vec{He_{\pm}})$ we use the  modified Glauber method 
\cite{bauer} 
which
takes into account the fact that the longitudinal 
momentum transferred in the transition $\gamma \rightarrow h$ is 
$q_{\parallel}=\frac{Q^2+M_{h}^2}{2q_{0}}$.
Within the above approximation we have $q_{\parallel}=2m_Nx$.
If we include  all possible permutations of the nucleons, we can write
the modified profile function in the following form
\begin{eqnarray}
\Gamma(\vec{\rho},\vec{r_{1t}},\vec{r_{2t}},\vec{r_{3t}})&=&\Gamma_{n}
(\vec{\rho}-\vec{r_{1t}})+2\,\Gamma_{p}(\vec{\rho}-\vec{r_{1t}})-4\,
\Gamma_{n}(\vec{\rho}-\vec{r_{1t}})\Gamma_{p}(\vec{\rho}-\vec{r_{2t}})
\Theta(z_2-z_1)e^{iq_{\parallel}(z_1-z_2)} \nonumber\\
&-&2\,\Gamma_{p}(\vec{\rho}-\vec{r_{1t}})\Gamma_{p}(\vec{\rho}-
\vec{r_{2t}})\Theta(z_2-z_1)e^{iq_{\parallel}(z_1-z_2)} \nonumber\\
&+&6\,\Gamma_{n}(\vec{\rho}-\vec{r_{1t}})\Gamma_{p}(\vec{\rho}-
\vec{r_{2t}})\Gamma_{p}(\vec{\rho}-\vec{r_{3t}})\Theta(z_2-z_1) 
\Theta(z_3-z_2)e^{iq_{\parallel}(z_1-z_3)}.
\end{eqnarray}
In these estimates we have accounted only for elastic rescatterings of the 
state $|h\rangle$.
It is a reasonable approximation at moderate $Q^2$.
 The scattering amplitudes $f$ are related to $\Gamma(\vec{\rho})$ as
\begin{equation}
f_{hp(n)}(q)=\frac{ik}{2\pi}\int e^{-i\vec{\rho}\cdot\vec{q_{t}}}\Gamma_{p(n)}
(\vec{\rho})d^2\vec{\rho}
\end{equation}

The $^3$He wave function is taken in a simple form ($S$-state), which 
works well in the Glauber calculations of 
elastic $p\,^4 {\rm He}$ scattering \cite{Levin}:
$|\Psi|^2\propto\prod_{l=1}^{l=3}\exp(-\vec{r_{l}}^2/2\alpha)
\delta^3(\sum \vec{r_{l}})$.
So, only the neutron is polarized in this approximation.
The numerical value of the slope was fixed to reproduce the e.m. form factor 
of $^3$He: $\alpha$=27 GeV$^{-2}$.
Within the described above approximation the $t$ dependence 
of the amplitude $hN \rightarrow hN$ is the same as for the 
amplitude $\gamma^* + N \rightarrow h + N$. Hence on the 
basis of current experience, we write
\begin{equation}
f_{hp}(q_{t})=is\sigma_{p}e^{-\beta/2\,q_{t}^2}(1+i\eta)
\end{equation}
\begin{equation}
f_{hn}^{\pm}(q_{t})=is\sigma_{n}^{\pm}e^{-\beta/2\,q_{t}^2}(1+i\eta_{\pm}),
\end{equation}
where $\eta=Re\,f_{hp}/Im\,f_{hp}$, 
$\eta_{\pm}=Re\,f^{\pm}_{hn}/Im\,f^{\pm}_{hn}$, 
 $\beta \approx 6$ GeV$^{-2}$. Note that since we 
are concerned here 
with the $x$ and $Q^2$ ranges corresponding to the  energies relevant to 
the current
measurements of $g_{1n}$, in
estimating $\beta$ from the HERA data we take into account a 
weak
energy dependence in the slope expected for the Regge 
pole approximation. We also assume that the slope for the spin 
dependent amplitude is the same as for the spin independent amplitude. Since 
both slopes are much smaller than the nuclear form factor slope our 
result is not sensitive to the value of $\beta$.
Finally we obtain for the total cross section 
\begin{eqnarray}  
\sigma^{\pm}_{T}&=&\sigma^{\pm}_{n}+2\sigma_{p}
-\frac{\sigma_{p}^2e^{-\alpha\, q_{\parallel}^2}}
{8\pi(\alpha+\beta)}
\Big(1-\eta^2-2\eta 
\sqrt{\frac{4\alpha}{\pi}}\cdot q_{\parallel} \Big) \nonumber\\
&-&\frac{\sigma^{\pm}_{n}\sigma_{p}e^{-\alpha\, 
q_{\parallel}^2}}{4\pi(\alpha+\beta)}\Big(1-\eta_{\pm}\eta-(\eta_{\pm}+\eta) 
\sqrt{\frac{4\alpha}{\pi}}\cdot q_{\parallel} \Big)+\frac{1}{48\pi^2
(\alpha+\beta)^2}\sigma_{p}^2\sigma^{\pm}_{n}
e^{-\alpha\, q_{\parallel}^2}.
\label{eq13}
\end{eqnarray}
In the third term, which is numerically small, we neglected the corrections 
due to the real part of the amplitude and higher order corrections in 
$q_{\parallel}$.
Using eq.(\ref{eq13}) we evaluate
the shadowing in the case of the unpolarized target:
\begin{eqnarray}
{F_{2A=3}(x,Q^2_0) \over3F_{2N}(x,Q^2_0)}=1-\frac{\sigma_{eff}}{8\pi
(\alpha+\beta)}exp(-\alpha q_{\|}^2)
\Big(1-\eta^2-2\eta 
\sqrt{\frac{4\alpha}{\pi}}\cdot q_{\parallel} \Big) \nonumber\\
+\frac{\sigma_{eff}^2}
{144\pi^2(\alpha+\beta)^2}\exp(-\alpha q_{\|}^2) . 
\label{f2}
\end{eqnarray}
For $g_{1^3He}(x,Q^2)$ we obtain:
\begin{eqnarray}
{g_{1^3He}(x,Q_0^2) \over g_{1n}(x,Q_0^2)}&=&
\frac{\sigma_{T}^{+}(e^3He)-\sigma_{T}^{-}(e^3He)}
{\sigma_{T}^{+}(en)-\sigma_{T}^{-}(en)}=\nonumber\\
&1&-\frac{\sigma_{p}exp(-\alpha q_{\|}^2)}{4\pi(\alpha+\beta)}
(1-K)+\frac{\sigma_{p}^2exp(-\alpha q_{\|}^2)}{48\pi^2(\alpha+\beta)^2} . 
\label{g1}
\end{eqnarray}
Here $K$ is given by
\begin{equation}
K=(\sqrt{\frac{4\alpha}{\pi}}q_{\parallel}+\eta) 
\cdot \frac{\sigma_{n}^{+}\eta_{+}-\sigma_{n}^{-}\eta_{-}}
{\sigma_{n}^{+}-\sigma_{n}^{-}} .
\end{equation}
Similar expressions are valid for the ratios~~ $g_{1^3H}/g_{1p}$ and 
$g^{n.s}_{1A=3}/g^{n.s.}_{1N}$.
Factors $\eta,~
 \eta_{\pm}$ are small 
because the vacuum exchange dominates in 
rescattering amplitudes.
This approximation may become dangerous for $x \le 10^{-3}$ where
$F_{2p}(x,Q^2)$ starts to increase fast with decrease of $x$.
At the same time the factor 
$(\sigma_{n}^{+}\eta_{+}-\sigma_{n}^{-}\eta_{-})/(\sigma_{n}^{+}-
\sigma_{n}^{-})$, which is determined by the phase of the secondary Regge
 trajectories which dominate 
in $g_{1N}(x,Q^2)$ for 
small $x$,  could be of  order  unity. 
However, its contribution
is suppressed for small $x$ by the
factor of $q_{\parallel}$
and the small value of $\eta$ for $x \ge 10^{-3}$.
 Hence in our estimates we neglect 
the contributions of the real part of all
amplitudes.
Uncertainties of this approximation will be analyzed elsewhere.

One can see from the comparison of eqs.(\ref{f2}) and (\ref{g1}) that 
shadowing for the case of the polarized cross section is larger by 
approximately a factor of two. This result justifies the above
qualitative discussion. Eq.(\ref{g1}) leads to  
${g_{1^3He}(x,Q_0^2) \over g_{1n}(x,Q_0^2)} \approx 0.9$  for $x \le 0.03$. 
Obviously, nuclear shadowing changes the contribution at small $x$ to the 
 Bjorken sum rule. As in the case for valence quarks (baryon sum rule)
and gluon distributions (momentum sum rule) the compensating positive 
 contribution to $g_{1A=3}^{n.s.}(x,Q^2)$ 
related to the projectile interaction with two nucleons
should be located at $x \sim 0.1$, cf. discussion
in \cite{FS88}. Hence we model this enhancement by requiring that
(i) the 
positive contribution to $g_{1A=3}^{n.s.}(x,Q^2)$
compensates the contribution due to shadowing in 
$\int_0^1 g_{1A=3}^{n.s.}(x,Q^2) dx$, (ii)  does not affect 
the region where shadowing 
is saturated
($x \le 0.03$), (iii)  it is concentrated for $x \le 0.15$. 
An example of this
fit is given in Fig.1 by a dashed line. One can see that 
typically the resulting enhancement is of the order $10 \div 15 \%$.

Thus we conclude that there are  two new effects modifying the   picture of 
nuclear effects for $g_{1^3\rm{He}}(x,Q^2)$ based on the 
nonrelativistic model of the nucleus: the  nonnucleonic 
degrees of freedom and nuclear shadowing. Based on the additive quark model 
one expects that in the high-energy limit $\sigma_{tot}(\left|h\right>\Delta)
\approx \sigma_{tot}(\left|h\right>N)$. Hence shadowing effect should be 
approximately the same for the nondiagonal contribution. 
Difference in the shadowing for higher partial waves maybe somewhat larger due
to smaller radius of these components of the wave function. However since the 
shadowing effect is rather small this effect does not exceed the product
of the shadowing correction and depolarization effect. Thus it leads to
an uncertainty in the discussed ratio of less than   1\%.
Hence the
two  discussed effects in the first approximation contribute multiplicatively 
to the modification of $g_{1A=3}^{n.s.}$ which is given by the solid curve.
Our expectations for the $g_{1^3He}/g_{1n}$ and $g_{1^3H}/g_{1p}$
ratios are practically the same.
It is noticeably  
 different from the $\sim 8\%$ depolarization effect obtained in 
the model \cite{Ciofi,Sauer} (the dashed-dotted line), where 
these effects were neglected.
Substantial model  dependence of the nuclear effects  
introduces significant uncertainties in the
extraction of $g_{1n}$ from the $^3$He data, especially for $x \le 0.2$.
The detailed procedure of extraction of $g_{1n}$ would involve 
separate modeling the
$\Delta$-contribution to $g_{1^3 \rm{He}}$ and $g_{1^3 \rm{H}}$ going beyond
 the $SU(6)$ -symmetry approximation and calculation of shadowing effects 
including effects of higher partial waves. 
We will consider these effects elsewhere.

In this  $x$-range $^2$H targets may have certain advantages since in this case
 nonnucleonic admixtures are much smaller due to weaker binding and zero 
isospin. The 
 shadowing effects are also smaller  for $g_{1^2\rm{H}}$ by a factor of
$\sim 0.4 \div 0.5 $ \cite{BBB}.  Besides, in 
the first approximation 
$g_{1p}(x,Q^2) \approx -g_{1n}(x,Q^2)$ for small $x$ in which case  
shadowing does not affect the extraction of $g_{1n}$.

Further studies are necessary to work out the  $x$-dependence of the
 contribution of the  nonnucleonic degrees of freedom and to develop a 
dynamic mechanism of the enhancement effects for $g_{1A=3}^{n.s.}$.
 
Also it would be interesting to check the predicted patterns for the 
screening-enhancement in independent experiments with other polarized 
nuclei where the polarization is carried  predominantly by a proton. 
Obviously, 
the heavier the nucleus, the larger the effect.
Another interesting question is the relation of shadowing to quenching
of $G_{A}/G_{V}$ for heavy nuclei. If there were no enhancement at 
moderate $x$ associated with shadowing at small $x$, it would lead to 
renormalization of $G_{A}/G_{V}$ by 10-20 \%.

We would like to thank 
N.Auerbach, T.Ericson,
G. Garvey, J. Friar, M. Karliner, G.Miller and P.Sauer for useful discussions.
We  are also indebted to A.Yu. Umnikov who after
our paper was released \cite{hep}
 has drawn our attention to their publication.

This work is supported in part by the U.S. Department of Energy and BSF.
One of us (M.S.)  thanks the DOE's Institute for Nuclear Theory at
the University of Washington for its hospitality and support during 
the workshop ``Quark and Gluon Structure of Nucleons and Nuclei''.

\begin{figure}
\centerline{
\epsfig{file=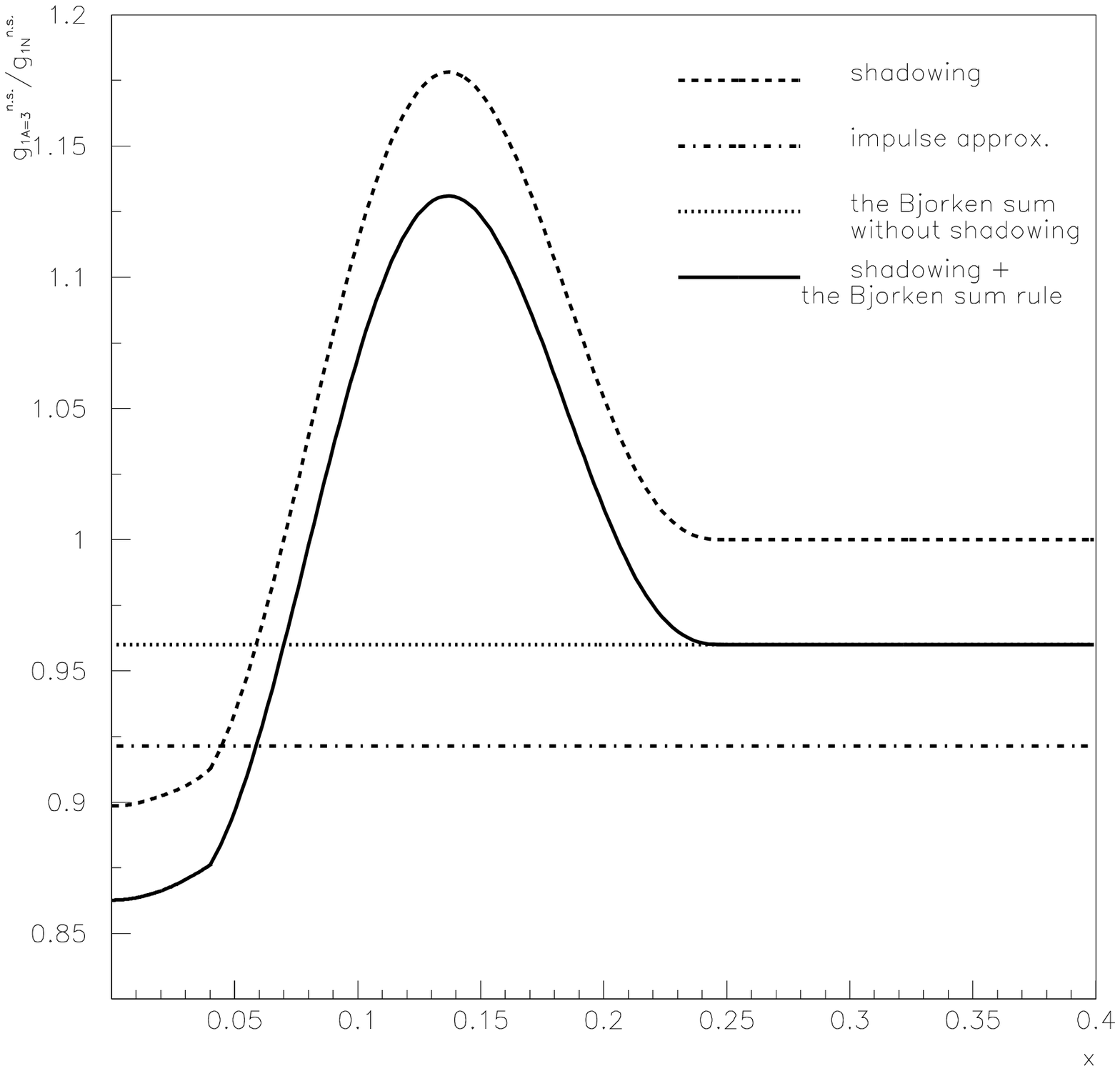,height=15cm,width=15cm}}
\caption{$g_{1A=3}^{n.s.}/g_{1N}^{n.s.}$ 
as a function of x. The dashed 
 line represents nuclear shadowing at small x. The solid line is 
the result of the fit constrained to preserve the Bjorken sum rule.}
\label{Fig.1}
\end{figure}
\end{document}